\documentstyle[aps,prb,multicol,epsfig,times]{revtex}
\makeatletter
\newlength\abovecaptionskip \newlength\belowcaptionskip
\def\@makecaption#1#2{%
 \vskip\abovecaptionskip \sbox\@tempboxa{#1: #2}%
 \ifdim \wd\@tempboxa >\hsize #1: #2\par \else \global \@minipagefalse
 \hb@xt@\hsize{\hfil\box\@tempboxa\hfil}%
 \fi \vskip\belowcaptionskip} \makeatother
\begin{document}
\title{Sublocalization, Superlocalization, and \\ Violation of Standard
Single Parameter Scaling in the Anderson Model}
\author{Jan W. Kantelhardt$^{1,2}$ and Armin Bunde$^1$}
\address{$^1$ Institut f\"ur Theoretische Physik III, 
Justus-Liebig-Universit\"at Giessen, D-35392 Giessen, Germany}
\address{$^2$ Center for Polymer Studies and Department of Physics, 
Boston University, Boston, MA 02215, USA}
\date{submitted: January 19, 2002, revised version: April 24, 2002}
\draft\maketitle\begin{multicols}{2}[%
\begin{abstract}
We discuss the localization behavior of localized electronic wave 
functions in the one- and two-dimensional tight-binding Anderson model
with diagonal disorder.  We find that the distributions of the local
wave function amplitudes at fixed distances from the localization
center are well approximated by log-normal fits which become exact at
large distances.  These fits are consistent with the standard single
parameter scaling theory for the Anderson model in 1d, but they suggest
that a second parameter is required to describe the scaling behavior
of the amplitude fluctuations in 2d.  From the log-normal distributions
we calculate analytically the decay of the mean wave functions.  For
short distances from the localization center we find stretched
exponential localization (``sublocalization'') in both, $d=1$ and $d=2$.
In $d=1$, for large distances, the mean wave functions depend on the
number of configurations $N$ used in the averaging procedure and decay
faster that exponentially (``superlocalization'') converging to simple
exponential behavior only in the asymptotic limit.  In $d=2$, in contrast,
the localization length increases {\it logarithmically} with the distance
from the localization center and sublocalization occurs also in the second
regime.  The $N$-dependence of the mean wave functions is weak.  The
analytical result agrees remarkably well with the numerical calculations.
\end{abstract}
]

\section{Introduction}

The Anderson model is the standard model for the electronic
properties of disordered solids,\cite{anderson-58} for reviews,
see e.~g. Refs.\cite{economou-90,kramer-93,janssen-98}  The wave
functions are known to be localized in dimensions $d \le 2$
(except at the band center for non-diagonal disorder) as well
as for sufficiently strong disorder in $d>2$.  The shape of these 
localized wave functions is believed to be characterized by an
asymptotically exponential decay, which can be described most
effectively\cite{anderson-80} by the Lyapunov exponent $\gamma$
or its inverse, the localization length $\lambda$.  The simple
exponential shape has been the basis of many calculations 
and models, e.~g. Mott's hopping conductivity and tunneling on 
quantum dots.  Even for other models of disordered matter, e.~g. 
for percolation models,\cite{bundestauffer} simple exponential 
localization of the wave functions has been assumed based on the 
analogy with the Anderson model.

In addition to the exponential decay at large distances from the
localization center, the wave functions in the Anderson model
exhibit large fluctuations of their amplitudes.\cite{kramer-93}
These fluctuations are rather important for
transport properties, e.~g. for the distribution of mesoscopic 
conductances, since the core region of the wave functions 
is dominated by the fluctuations, and the exponential decay can 
only be observed at distances much larger than the localization 
length $\lambda$ for single wave functions.  In the single-parameter 
scaling theory\cite{abrahams-79,anderson-80} (SPST) it has been
assumed that the whole distribution of the conductances or Lyapunov
exponents for an ensemble of configurations can be described by one 
parameter only, e.~g. the width of the distribution of values of 
Lyapunov exponents depends on the average Lyapunov exponent $\gamma$ 
in a universal way.  Hence, the localization length $\lambda = 
1/\gamma$ could be used as single relevant scaling parameter 
fully characterizing the localization behavior.  

Very recently, there has been a growing interest in the question, 
if the behavior of the fluctuations of the wave functions are 
universal and can be described within the
SPST,\cite{deych-00,braun-01,ruehlaender-01} see also Refs.
\cite{anderson-80,mueller-97,uski-98,deych-98,uski-01,nikolic-01}
The interest is partly motivated by the experimental observations of 
a transition in the behavior of the conductance of low density two 
dimensional semiconductor devices from an insulating like temperature 
dependence at low densities to a metallic one at higher
densities.\cite{abrahams-01}  This apparent localization-delocalization
transition in $d=2$ (2d) is at odd with the SPST for non-interacting
electrons.

In this paper, we investigate the distribution of the amplitudes of
Anderson localized wave function in $d=1$ (1d) and 2d at given
distances $r$ from the localization center, which allows a test of 
the SPST hypothesis.  We show that the SPST hypothesis holds 
perfectly in 1d, while deviations from the standard SPST occur
for localized wave functions for the Anderson model in 2d.  
Deviations also occur for the percolation model in 2d and 3d.
In both cases, the width of the amplitude distribution is characterized 
by a second parameter that depends on the type of disorder.  In addition, 
we find a clear {\rm logarithmic} system-size dependence of the
localization length $\lambda$ for the Anderson model in 2d, but no
significant dependence on the boundary conditions.

From the amplitude distributions we can calculate analytically the 
behavior of the mean wave functions at sufficiently large distances 
from the localization center.  We find that simple exponential 
decay as usually anticipated occurs only in 1d in the asymptotic limit
where the wave function has decayed to extremely small values.  At 
finite distances from the localization center, we observe two 
different localization regimes.  In the first regime, at short 
distances from the localization center, we find stretched exponential 
localization with a localization exponent below one (``sublocalization'')
in 1d and 2d.  In the second regime, at larger distances, the averages 
in 1d depend on the number of configurations and we observe 
``superlocalization'', where the localization exponent is greater than 
one.  The crossover from the sublocalization to the superlocalization 
regime increases logarithmically with the number of configurations.
In the asymptotic limit, the localization exponent converges to one, 
yielding the theoretically predicted simple exponential behavior.  
In $d=2$, however, we observe the surprising result that the 
sublocalization of the mean wave functions remains also in the second 
regime.  

The paper is organized as follows:  In Section II we briefly introduce
the Anderson model and summarize some of the known results.  In 
Section III we discuss distributions of wave function amplitudes, 
beginning with a review of the predictions of the standard SPST and 
the random matrix theory.  Then we introduce an amplitude statistics
which is most useful for the characterization of localized wave
functions, and present our numerical results for this amplitude 
statistics.  In Section IV we calculate analytically the decay of the
mean wave function amplitudes and distinguish different localization
regimes.  After describing an appropriate averaging procedure, we
present our numerical results for the decay of the wave functions in
Section V.  We summarize our results in Section VI.

\section{The Anderson Model}

We consider the Schr\"odinger equation in tight-binding approximation
for the wave function of a quantum particle on regular lattices.  The
coefficients $\psi_n$ for each lattice site $n$ satisfy the
tight-binding equation
\begin{equation}
 E \, \psi_{n} = \epsilon_n \, \psi_{n} + \sum_\delta V_{n,n+\delta} 
 \, \psi_{n+\delta}, \label{eq:TB} \end{equation}
where the sum runs over all nearest neighbor sites $n+\delta$ of site
$n$. $\vert\psi_{n}\vert^2$ is the probability density on site $n$
for the wave function with energy $E$.  The hopping terms
$V_{n,n+\delta}$ are constant for nearest neighbor cluster sites and
zero otherwise; for simplicity we take $V=1$ as energy unit.  The
values for the diagonal terms $\epsilon_n$ are randomly chosen from
the interval $[-w/2, w/2]$, where $w$ is the disorder parameter
(diagonal disorder).  

Here, we consider the Anderson model with periodic as well as hard wall 
boundary conditions for a linear chain and on the square lattice.  
In these cases the wave functions are known to be
localized\cite{economou-90,kramer-93,abrahams-79,mackinnon-81} even
for small disorder strength $w$.  For large distances $r$ from the
localization center, the decay of the eigenfunction amplitudes is
believed to be exponential.  Thus, a localization length $\lambda$
and the Lyapunov exponent $\gamma \equiv \lambda^{-1}$ can be
defined to describe the asymptotic decay of the mean logarithm of
the amplitudes,
\begin{equation}
\left\langle \ln \vert \psi(r) / \psi(0) \vert \right\rangle = 
- r/\lambda = -r \gamma \label{eq:lokl} \end{equation}
for $r \gg \lambda$.  In 1d the localization length in the band center 
$E \simeq 0$ can be calculated analytically from a recursion equation 
derived from Eq.~(\ref{eq:TB}) using Greens functions:\cite{kappus-81} 
$\lambda(w) \simeq 105.045 /w^2$.  This result has also been verified
numerically.\cite{pichard-86}  While the prefactor slightly depends 
on the energy $E$, the $1/w^2$-dependence holds for all eigenstates
except at the band edges, where a different analytical theory
applies.\cite{russ-98}  In 2d there is no analytical formula for the 
localization length $\lambda$.  The numerical results published so far
were obtained using finite size scaling combined with (a) transfer matrix
calculations of the conductance\cite{mackinnon-81} and (b) exact
diagonalization and level statistics.\cite{zharekeshev-96}

We are interested in the distribution of the amplitudes $\vert \psi_n
\vert$ for the eigenfunctions of Eq.~(\ref{eq:TB}).  Because of the 
disorder, the eigenvalue equation has to be solved numerically.  
In 1d we use an iteration method,\cite{roman-87} while we employ a
Lanczos algorithm\cite{lanczos} in 2d.

\section{The Amplitude Distributions}

\subsection{Definition}

In the original single-parameter scaling theory\cite{abrahams-79} 
(SPST) the fluctuations of the wave functions had not been considered.
An extension of the theory includes the scaling of the
fluctuations.\cite{anderson-80}  Still it has been assumed that there
is only one relevant scaling parameter characterizing the localization
behavior {\it and} the fluctuation behavior.  This means that, e.~g.,
the position of the center as well as the width of the distributions
of conductances, Lyapunov exponents, or localization lengths can be
characterized by only one scaling variable.  Practically the
distribution of the parameter
\begin{equation} \tilde\gamma (L) = {1 \over 2L} \ln (1+1/g) \label{gamma}
\end{equation}
with the conductance $g$ and the system size $L$ has been investigated
theoretically\cite{anderson-80} within the so-called random-phase
hypothesis, see Ref.\cite{deych-00} for a thorough discussion.  While the
average $\langle \tilde\gamma(L) \rangle$ was shown\cite{mackinnon-81} to
approach the Lyapunov exponent $\gamma$,
\begin{equation} \langle \tilde\gamma(L) \rangle = \gamma \equiv
\lambda^{-1} \qquad {\rm for} \quad L \gg \gamma^{-1}, \label{eq:mi1}
\end{equation}
its fluctuations were expected to become
\begin{equation} \langle \tilde\gamma^2(L) \rangle - \langle
\tilde\gamma(L) \rangle^2 = \gamma /L \qquad {\rm for} \quad
L \gg \gamma^{-1} \label{eq:flu1} \end{equation}
asymptotically.  Thus, $\gamma$ fully characterizes the asymptotic
scaling behavior of the average $\tilde\gamma(L)$ {\it and} its
fluctuations.  Although many numerical calculations of $\langle \tilde
\gamma(L) \rangle$ [testing Eq.~(\ref{eq:mi1})] have been reported,
there is only little work devoted to the scaling behavior of the
fluctuations and their relation to $\gamma$.  Recently Deych et 
al.\cite{deych-00} considered the Lloyd model (which is identical to the
Anderson model with diagonal disorder except for a Cauchy distribution of
the site energies) and found that states at the tail of the density of
states violate the SPST and that there exists a new length scale that is
responsible for this violation.

In this paper, we determine numerically the distribution of the
logarithm of normalized amplitudes, $A(r) \equiv -\ln \vert \psi(r)
/\psi(0) \vert$, for large distances $r$ from the localization
center (at $r=0$).  The distance $r$ is directly related to the system
size $L$ used in Eqs.~(\ref{gamma})-(\ref{eq:flu1}), since the value and
the fluctuations of the conductance $g$ for a system of size $L$
reflect the value and the fluctuations of its wave functions at the
distance $r=L$ from the localization center.  Following
Eqs.~(\ref{eq:lokl}) and (\ref{eq:mi1}), it is obvious that
\begin{equation} \langle A(r) \rangle \equiv \left\langle -\ln \left\vert
{\psi(r) \over \psi(0)} \right\vert \right\rangle = {r \over \lambda}
= r \langle \tilde\gamma(r) \rangle \label{eq:Amit} \end{equation}
for $r \gg \lambda$.  Hence, $A(r)$ corresponds to $r \tilde\gamma(r)$,
and the fluctuations of $A(r)$ are expected to become 
\begin{equation} \langle A^2(r) \rangle - \langle A(r) \rangle^2
= (\langle \tilde\gamma^2 \rangle - \langle \tilde\gamma \rangle^2) 
r^2 = \gamma r = r / \lambda  \label{eq:flu2} \end{equation}
for large $r$ in the standard SPST with the random-phase hypothesis 
following Eq.~(\ref{eq:flu1}).  In order to estimate possible
deviations from this standard single parameter scaling behavior, we
will calculate the fluctuation parameter
\begin{equation} \sigma(r) \equiv 2 \left[\langle A^2(r) \rangle -
\langle A(r) \rangle^2\right] \lambda(r) / r, \label{eq:sigma}
\end{equation}
which is expected to become $\sigma = 2$ asymptotically in the
standard SPST\cite{anderson-80} according to Eq.~(\ref{eq:flu2}) for the
Anderson model and $\sigma = 4$ for the Lloyd model.\cite{deych-00} 

In an alternative approach it has been shown some years ago, that 
the random matrix theory\cite{metha-91} can be applied to
electronic eigenfunctions in disordered systems in
3d\cite{shklovskii-93} and in 2d.\cite{zharekeshev-96}  In 2d,
for small disorder strength $w$, the localization length becomes 
larger than the system size, $\lambda \gg L$, and the wave functions 
become quasi extended. In this regime, as well as in the metallic 
regime in 3d, the random matrix theory correctly describes the level
spacing distribution as well as the distribution of the eigenfunction
amplitudes $\vert \psi_n \vert$, which are usually assumed to be 
normalized according to $\sum_n \vert \psi_n \vert^2 = 1$.  The 
intensity distribution (histogram) is usually defined as 
\cite{mueller-97,uski-98,uski-01,nikolic-01,falko-96}
\begin{equation} f(t) = \left\langle {1 \over N} \sum_{n=1}^N 
 \delta( t - \vert \psi_n^{(E')} \vert^2 N) \right\rangle_E,
\label{eq:ft} \end{equation}
where the sum runs over all lattice sites, $N$ is the total
number of lattice sites, and the average $\langle \cdots \rangle_E$ 
is taken over several eigenfunctions $\psi_n^{(E')}$ with energy
$E' \approx E$.  In the limit of small disorder strength $w$,
$f(t)$ is given by the universal Porter-Thomas 
distribution,\cite{porter-56}
\begin{equation} f_{\rm PT}(t) = {1 \over \sqrt{2 \pi t}} 
\exp (-t/2). \end{equation}
This result has been confirmed numerically recently for quasi extended
states in the Anderson model in 2d\cite{mueller-97} and for
the metallic regime in 3d.\cite{uski-98,nikolic-01}

The intensity distribution $f(t)$, however, cannot appropriately 
describe the behavior of localized wave functions in the limit of 
strong localization $\lambda \ll L$.  The analytical form of $f(t)$ 
obtained by inserting a simple exponential decay without any amplitude 
fluctuations into the definition (\ref{eq:ft}) was shown to be 
consistent with the numerical results for real fluctuating Anderson 
wave functions.\cite{mueller-97}  Thus, a simple exponential decay 
without any fluctuations cannot be distinguished from a real fluctuating 
wave function, if (\ref{eq:ft}) is used to calculate the intensity 
statistics.  The fluctuations simply do not affect $f(t)$.  This 
indicates that the definition (\ref{eq:ft}) is not appropriate to 
fully characterize localized eigenfunctions.  Further, since no reference 
to the localization center is taken into account in (\ref{eq:ft}), the
calculation of the decay of mean wave functions is not possible from
$f(t)$.  Another problem is that $f(t)$ depends on the system size $N$ 
for localized states.  So, we introduce a different definition for the 
amplitude distribution here, which avoids these problems.

Since we are interested in the decay of the localized wave functions,
we investigate the distribution of the amplitudes $\vert \psi \vert$
at a certain distance $r$ from the localization center.  By definition, 
the localization center for each eigenfunction $(\psi_n^{(E)})$ is that
site $n_0$, where the maximum of the amplitudes occurs, i.~e.
$\vert \psi_{n_0}^{(E)} \vert = \max_n \vert \psi_n^{(E)} \vert$.
An alternative definition for the localization center would be to 
consider the lattice point most close to the ``center of gravity'' of 
the wave function as the center of localization.  That is the site $n_1$ 
where $\sum_n [(x_n-x_{n_1})^2 + (y_n-y_{n_1})^2] \vert \psi_n \vert^2$ 
reaches its minimum value ($x_n$ and $y_n$ designate the coordinates of
site $n$ here).  Often, these two definitions of the localization
center coincide.  But even if they do not coincide, there is
practically no difference in the results for the amplitude
distributions, as we will show below.  In order to achieve a system-size
independent normalization of the amplitudes $\vert\psi\vert$, which
allows us to compare numerical data for different system sizes and
simplifies the analytical description significantly, we choose to set
$\vert \psi_{n_0} \vert = 1$.  Note that this choice is different from the
usual normalization $\sum_n \vert \psi_n \vert^2 = 1$ underlying the
definition (\ref{eq:ft}), since the sum is always larger than one for
our normalization. 

To describe the actual decay of the localized wave functions, we do not 
consider the distribution of the amplitudes for the whole eigenfunction 
[as in the definition (\ref{eq:ft})], but instead look at the distributions 
of the amplitudes at given distances $r$ from the localization center.  
Hence, we replace $f(t)$ by the histogram distribution function $H(A, r)$ 
defined by
\begin{equation}
 H(A,r) = \left\langle {1 \over N_r} \sum_{n=1}^N \delta(r - r_n)
 \delta( A + \ln \vert \psi_n^{(E')} \vert) \right\rangle_E,
\label{eq:HAr} \end{equation}
where $r_n$ is the distance of the site $n$ from the localization
center, and $N_r$ is the number of sites with $r_n = r$, i.~e. 
$N_r = \sum_{n=1}^N \delta(r - r_n)$.  In the definition (\ref{eq:HAr}), 
we use the variable $A \equiv -\ln \vert \psi_n^{(E')} \vert$ instead
of $t \equiv \vert \psi_n^{(E')} \vert^2 N$ from (\ref{eq:ft}) for two
reasons:  (i) $t$ is system-size dependent for localized states, and
(ii) $A$ can be related much more easily to the predictions of the
standard SPST, that we want to test.  In fact, we have already
derived the relation in Eqs.~(\ref{eq:Amit}) and (\ref{eq:flu2}).
Note that $H(-\ln \vert \psi \vert, r)$ is the probability density of 
finding the amplitude $\vert \psi \vert$ at distance $r$ from the 
localization center.  In the following, we will use this definition 
of the eigenfunction amplitude distributions to present our numerical
results and discuss the localization behavior.

\begin{figure} \centering
\epsfxsize8.cm\epsfbox{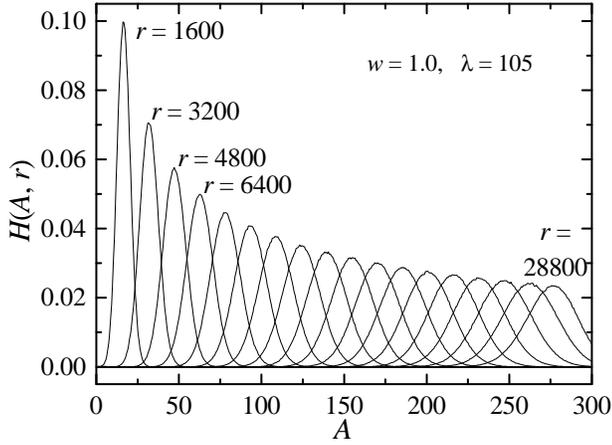}
\parbox{8.5cm}{\caption[]{\small
Histogram distributions $H(A,r)$ of the absolute amplitude values
$\vert\psi\vert$ at distance $r$ from the localization center versus
$A \equiv -\ln \vert\psi\vert$ for electronic wave functions in the
Anderson model in 1d with disorder strength $w = 1.0$ at several fixed
distances $r$ from the center of localization ranging from $r=1600$
(leftmost curve) with increment $\Delta r = 1600$ to $r=28800$
(rightmost curve).  To determine the distributions, more than $10^6$
eigenfunctions have been calculated on chains of lengths $L=30000$.}
\label{fig:1}} \end{figure}

\subsection{Results in 1d}

\begin{figure} \centering
\epsfxsize8.cm\epsfbox{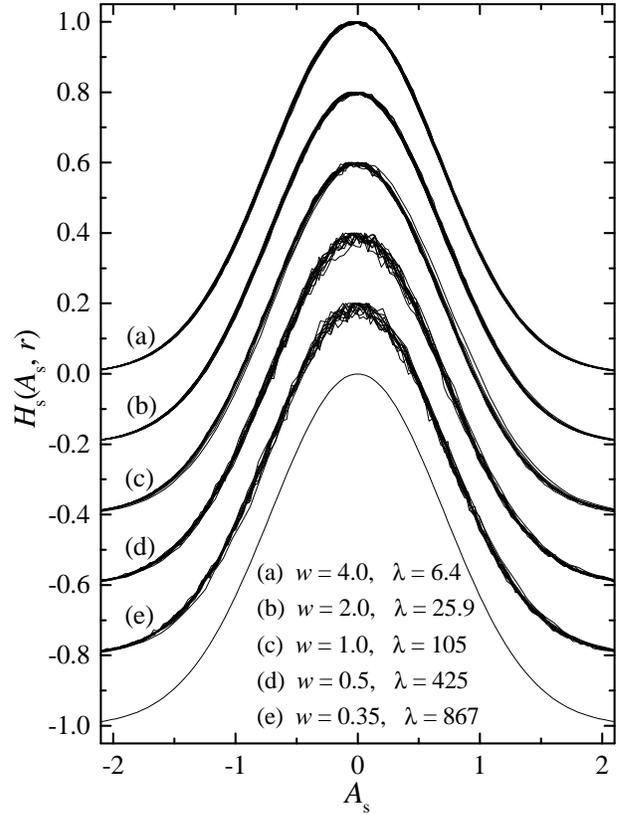}
\parbox{8.5cm}{\caption[]{\small
Scaled distributions $H_{\rm s}(A_{\rm s},r) \equiv H(A,r)
\sqrt{\pi\sigma r/\lambda}$ of the amplitude values $\vert\psi\vert$ at
distance $r$ from the localization center versus $A_{\rm s} \equiv
(A - r/\lambda) / \sqrt{\sigma r /\lambda}$ for electronic wave functions 
in the Anderson model in 1d with disorder strength $w = 4.0$, 2.0, 1.0, 
0.5, and 0.35 at several distances $r$ from the center of localization 
ranging from $r \approx 20 \lambda$ to $250 \lambda$.  Due to the scaling 
the data collapse onto single curves.  The curves for different $w$ have 
been shifted by multiples of 0.2.  The continuous line below the data 
corresponds to the ansatz Eq.~\protect(\ref{eq:pvonA}).  To determine the 
distributions, (a) 4,200,000, (b) 3,400,000, (c) 1,100,000, (d) 200,000
and (e) 170,000 eigenfunctions have been calculated on chains of lengths 
$L > 250 \lambda(w)$.}
\label{fig:2}} \end{figure}

Figures~\ref{fig:1} and \ref{fig:2} show our numerical results of the 
amplitude distribution functions $H(-\ln \vert \psi \vert, r)$ for the 
Anderson model in 1d with periodic boundary conditions.  Using an
iteration method\cite{roman-87} we have calculated eigenfunctions
with energy $E \approx 0$ (corresponding to the band center) for linear
chains with five different disorder strengths $w$ and determined the
distribution functions $H(A,r)$ for several values of $r$.  
Figure~\ref{fig:1} shows 18 of these histogram distributions calculated 
for more than $10^6$ chains of length $L=30000$ with $w=1.0$.

We find that the histograms $H(-\ln \vert \psi \vert, r)$ of the
amplitudes $\vert \psi \vert$ at fixed distances $r$ from the 
localization center surprisingly well follow a log-normal ansatz,
\begin{equation}
 H_{\rm ln}(-\ln \vert \psi \vert, r) \equiv \sqrt{\lambda \over {\pi 
 \sigma \, r}} \, \exp \left[-{ (\ln \vert \psi \vert + r/\lambda)^2 
 \over \sigma \, r/\lambda}\right].
\label{eq:pvonA} \end{equation}
In general, both parameters, $\lambda$ and $\sigma$, might depend on
$r$.  Since $H_{\rm ln}(A,r)$ is a simple Gaussian
distribution with average value $\langle A(r) \rangle = \int A
H_{\rm ln}(A,r) \, dA = r/\lambda$ and variance $\langle A^2(r) \rangle - 
\langle A(r) \rangle^2 = \sigma r /2 \lambda$, the two parameters 
$\lambda$ and $\sigma$ correspond to the localization length and the 
fluctuation parameter, respectively, according to Eqs.~(\ref{eq:Amit}) 
and (\ref{eq:sigma}).  Note that the amplitude $\vert \psi \vert = 
\exp(-r/\lambda)$ occurs with the largest probability at distance $r$ 
from the center of localization, which further prompts us to identify 
the parameter $\lambda$ with the localization length.  In the following,
$\lambda(r)$ always refers to the parameter determined from the 
distribution $H(A,r)$ for a given $r$, while $\lambda$ represents the 
asymptotic value of $\lambda(r)$.

The log-normal distribution of the amplitudes can be motivated by the 
following argument.  If 1d wave functions are calculated using the 
transfer matrix method, the value of $\psi_n$ for large $n$ is related to 
the values of $\psi_n$ for the two initial sites, $n=0$ and $n=1$, by the 
product of $n$ transfer matrixes $M_m$,
\begin{equation} \label{matrix}
\left[ \prod_{m=n}^1 M_m \right] {\psi_1 \choose \psi_0} = 
{\psi_{n+1} \choose \psi_{n}}, \quad M_m = 
{ E-\epsilon_m \; -1  \choose  1 \qquad \quad 0 }. \end{equation}
This product of matrices will involve many products of the random site 
energies $\epsilon_n$.  Products of many independently distributed 
random numbers asymptotically follow a log-normal distribution 
according to the central limit theorem (under some conditions, which we 
do not want to discuss in detail here).  Thus, it is reasonable that
$\psi_n$ will also follow a log-normal distribution asymptotically for 
large $n$. 

The agreement between the numerical distributions $H(A,r)$ (shown in
Fig.~\ref{fig:1}) and the log-normal distribution $H_{\rm ln}$
[Eq.~(\ref{eq:pvonA})] becomes perfect for large $r$ values.  
This indicates that the distribution $H(A,r)$ follows 
Eq.~(\ref{eq:pvonA}) asymptotically.  In addition, we find that both 
parameters, $\lambda$ and $\sigma$, are independent of $r$.  Fits of 
Eq.~(\ref{eq:pvonA}) to the numerical distributions $H(A,r)$ for $w=1.0$ 
shown in Fig.~\ref{fig:1} give $\lambda = 105 \pm 1$ and $\sigma=2.1 
\pm 0.15$.  The value of $\lambda$ is in perfect agreement with the
analytical result\cite{kappus-81} $\lambda(w) \simeq 105.045 /w^2$.
The value of $\sigma$ is consistent with $\sigma=2$ expected in the 
standard SPST.  

Since both parameters do not depend on $r$, the distributions
shown in Fig.~\ref{fig:1} can be rescaled in order to obtain a data 
collapse for the different distances $r$ from the localization center.  
If we multiply $H(A,r)$ by $\sqrt{\pi\sigma r/\lambda}$ and replace 
$A$ by $A_{\rm s} \equiv (A - r/\lambda) / \sqrt{\sigma r /\lambda}$, 
all distribution will collapse upon each other.  This is shown in 
Fig.~\ref{fig:2}, where data for the other disorder strengths
$w$ we considered are also included.  Since the dependence of the 
localization length $\lambda$ on $w$ is known, and the fluctuation
parameter $\sigma$ turns out to be independent of $w$, we are able 
to rescale the distributions for all $r$ and $w$ in order to obtain 
a full data collapse also for the other disorder strengths. 
The numerical result $\sigma = 2.1 \pm 0.15$, which is consistent 
with $\sigma=2$, fully confirms the standard SPST for the Anderson 
model in 1d.

\subsection{Results in 2d}

For the Anderson model in 2d the scaling behavior of the histograms
of amplitudes $\vert \psi \vert$ at fixed (Euclidian) distance $r$ from 
the localization center turns out to be not as simple as in 1d.  Using
the Lanczos algorithm \cite{lanczos} we have calculated eigenfunctions 
of Eq.~(\ref{eq:TB}) for energies $E \approx 0$ and two disorder strengths
$w=8.5$ and $w=10.0$.  For these values of $w$ the amplitudes of the
localized states decay roughly to $10^{-13}$ and $10^{-20}$, respectively,
on our $300 \times 300$ square lattice.  This allows to observe the
localization effects over several orders of magnitude.  Quadruplicate
precision is used in all calculations in order to get reliable values
also for small amplitudes $\vert\psi\vert \approx 10^{-20}$.  We compare
the results for periodic boundary conditions (PBC) and hard wall boundary
conditions (HWBC).

\begin{figure} \centering
\epsfxsize8.cm\epsfbox{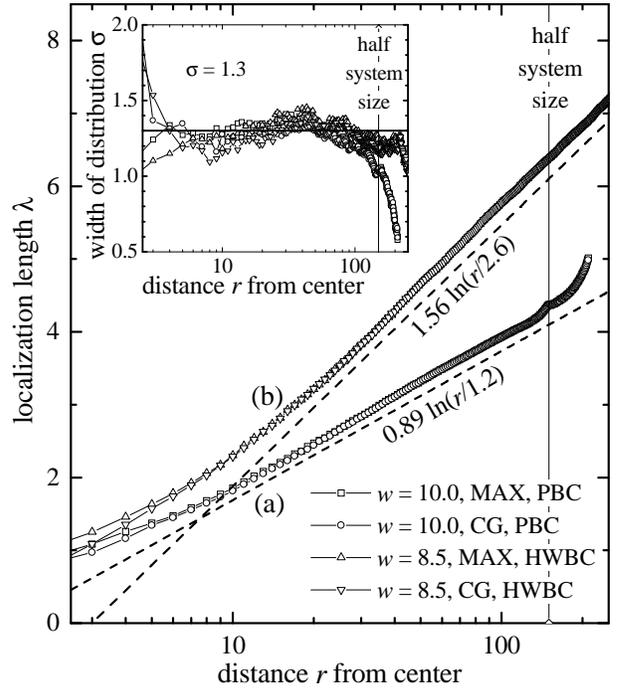}
\parbox{8.5cm}{\caption[]{\small
Localization length $\lambda(r)$ of Anderson wave functions in 2d
in the localized regime for (a) $w=10.0$ and (b) $w=8.5$.  The values
of $\lambda(r)$ have been determined by fits of Eq.~(\protect\ref{eq:pvonA})
to the amplitude distributions $H(A,r)$ for each distance $r$ from the
localization center.  Data for periodic boundary conditions (PBC, a)
as well as hard-wall boundary conditions (HWBC, b) have been analyzed.
The results for both definitions of the localization center [as the
maximum (MAX) and as the ``center of gravity'' (CG)] are practically
identical (except for $r<10$) and well described by a logarithmic
dependence of $\lambda$ on $r$:  (a) $\lambda(r)/0.89 = \ln(r/1.2)$ and
(b) $\lambda(r)/1.56 = \ln(r/2.6)$.  The straight lines correspond to
these fits, but they are shifted by a constant offset for visibility. 
In the inset, the estimated fluctuation parameters $\sigma(r)$ of the
distributions are shown.  They are approximately independent of $r$ and
$w$.  We find $\sigma = 1.3 \pm 0.2$.  For both disorder strengths, 
about 1000 eigenfunctions were calculated with the Lanczos algorithm
with quadruplicate precision on a $300\times300$ lattice.}
\label{fig:3}} \end{figure}

We find that the histogram distributions $H(A, r)$ can still be well
approximated by the log-normal distribution $H_{\rm ln}(A,r)$ 
[Eq.~(\ref{eq:pvonA})].  But now the parameter $\lambda(r)$ that 
corresponds to the localization length clearly depends on $r$.  This 
dependence is shown in the main part of Fig.~\ref{fig:3}, where $\lambda$ 
is plotted versus $r$ on a log-linear scale for systems with the disorder 
strength $w=10.0$ and PBC as well as for $w=8.5$ and HWBC.  We also 
checked $w=8.5$ with PBC and $w=10.0$ with HWBC and found no significant 
alterations in the behavior.  Further, the results do not change if we
disregard all those eigenfunctions that did not reach a preset accuracy 
limit in the Lanczos calculations or select only eigenfunctions within a
reduced energy interval.  The results also do not depend on the way the
center of localization is defined (as maximum or ``center of gravity'',
see Section III.A).

The values of $\lambda(r)$ and $\sigma(r)$ shown in Fig.~\ref{fig:3} have
been obtained by fitting Eq.~(\ref{eq:pvonA}) to $H(A, r)$ for each $r$.
Finite-size effects are clearly observable for PBC, while they seem to 
be rather negligible for HWBC.  The rather strong finite-size effects 
in the case of PBC are probably due to the many additional paths leading 
from the localization center to a remote site via the edges of the lattice.
In all cases the $\lambda(r)$ dependence is clearly a {\it logarithmic} one 
for $r>10$: 
\begin{equation} \label{lambdar} \lambda(r) / c_\lambda = \ln (r / c_r),
\end{equation}
where the two parameters $c_\lambda$ and $c_r$ describe the scales of the
localization length $\lambda$ and the distance $r$, respectively. 
Only for small $r$ values (i.~e. $r \ll 20$), a power-law fit is also
possible.  The logarithmic dependence is observed over more than one order
of magnitude in $r$ in all our numerical simulations. 

\begin{center}\begin{tabular}[t]{c|c|c|c|c|c|c}
$w$ & BC & $E_{\rm max}$ & number & $c_\lambda$ & $c_r$ & $\sigma$ \\ \hline
8.5 & HWBC & 0.05 & 994 & 1.56 &  2.6 & 1.35 \\
8.5 & HWBC & 0.005 & 359 & 1.54 & 2.5 & 1.34 \\
8.5 & HWBC & 0.001 & 76 & 1.53 & 2.4 & 1.28 \\ \hline
8.5 & PBC & 0.05 & 227 & 1.64 & 3.1 & 1.27 \\
8.5 & PBC & 0.005 & 73 & 1.68 & 3.2 & 1.37 \\ \hline
10.0 & HWBC & 0.05 & 288 & 0.91 & 1.3 & 1.32 \\
10.0 & HWBC & 0.005 & 97 & 0.92 & 1.4 & 1.34 \\ \hline
10.0 & PBC & 0.05 & 1029 & 0.89 & 1.2 & 1.25 \\
10.0 & PBC & 0.005 & 309 & 0.90 & 1.2 & 1.24 \\
10.0 & PBC & 0.001 & 68 & 0.92 & 1.4 & 1.20 \end{tabular}\end{center}

\noindent\parbox{8.5cm}{\small TABLE 1: The values of $c_\lambda$ and
$c_r$ from fits of Eq.~(\protect\ref{lambdar}) to the localization length
$\lambda(r)$ as well as the distribution width $\sigma$ for the Anderson
model in 2d.  The results for two disorder strengths $w=8.5$ and
$w=10.0$, for both types of boundary conditions (BC, HWBC = hard-wall,
PBC = periodic boundary conditions), and for different subsets of the
eigenfunctions are reported.  The value of $E_{\rm max}$ determines the
width of the energy interval $[0,E_{\rm max}]$ for the selected
eigenfunctions.  The number of eigenfunctions involved in each fit is
also reported.  The fitting range is $30 > r > 120$ in all cases, and
the maximum has been used to define the localization center.
\label{tab:1}}\vspace{0.5cm}

Table 1 reports the fitted values of $c_\lambda$ and $c_r$ for both
disorder strengths, $w=8.5$ and $w=10.0$, for both types of boundary
conditions, and for different subsets of the eigenfunctions.  The value of
$E_{\rm max}$ determines the width of the energy interval $[0, E_{\rm max}]$
for the selected eigenfunctions.  The table shows that $c_\lambda$ and $c_r$
clearly depend on $w$, but not on the width of the energy interval.  The
values for different boundary conditions are also very close to each
other, and we cannot find any systematic dependence upon the boundary
conditions.  Both, $c_\lambda$ and $c_r$, are rather close to the microscopic
distance between the sites (set to one here), but they are definitely not
simply identical to the microscopic distance, since both depend strongly on
the disorder parameter $w$.  The significant {\it logarithmic} dependence
of the localization length $\lambda(r)$ on the distance $r$ from the
localization center may explain the somewhat different numerical results
for the localization length that have been obtained 
before.\cite{mackinnon-81,zharekeshev-96}  When comparing with those results
one has to keep in mind, though, that they were determined using finite size
scaling of the conductance and level statistics, respectively, while our
determination does not involve finite size scaling.  In the finite size
scaling techniques, one value of the localization length $\lambda_L$ or a
similar level statistics parameter is calculated for each configuration. 
Then the average finite size parameters are extrapolated to infinite system
size by rescaling the results for different disorder $w$ in a scaling plot,
assuming the existence and scaling behavior of $\lambda_\infty$.  Here,
in contrast, we do not assume any scaling properties of the localization
length and determine the dependence of $\lambda(r)$ upon the distance $r$
from the origin for each system.

In contrast to $\lambda(r)$, the parameter $\sigma$ characterizing the
width of the amplitude distributions and shown in the inset of
Fig.~\ref{fig:3} is still independent of $r$, except for the pronounced
finite-size effects in the case of PBC.  We find $\sigma = 1.3 \pm 0.2$
for both disorder strengths $w$, both types of boundary conditions,
and all selections of eigenfunctions as reported in Table 1.  This value
clearly deviates from $\sigma = 2$ expected for the standard SPST.  

\begin{figure} \centering
\epsfxsize7.8cm\epsfbox{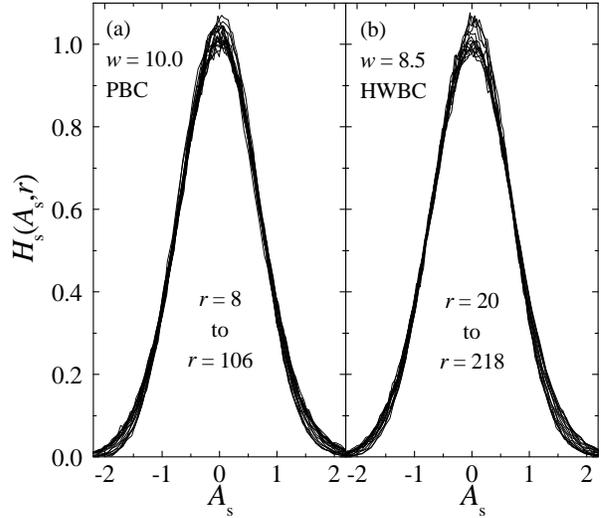}
\parbox{8.5cm}{\caption[]{\small
Scaled distributions $H_{\rm s}(A_{\rm s},r) = H(A,r)
\sqrt{\pi\sigma r/\lambda(r)}$ of the amplitude values $\vert\psi\vert$
at (Euclidian) distance $r$ from the localization center versus
$A_{\rm s} = [A - r/\lambda(r)] / \sqrt{\sigma r /\lambda(r)}$
for electronic wave functions in the Anderson model in 2d with
disorder strength (a) $w=10.0$ (PBC) and (b) $w=8.5$ (HWBC) at several
fixed distances $r$ from the center of localization, (a) $r = 8, \ldots,
106$, and (b) $r = 20, \ldots, 182$.  The $r$-dependent localization
lengths $\lambda(r)$ for the rescaling procedure have been taken from
logarithmic fits in Fig.~3, and the width parameter $\sigma$ is
fixed to 1.35.  For each part of the figure, $\approx 1000$ eigenfunctions
were calculated using the Lanczos algorithm with quadruplicate precision
on a $300\times300$ lattice. }
\label{fig:4}} \end{figure}

Using the logarithmic $\lambda(r)$ dependence, Eq.~(\ref{lambdar}), with
the parameters from Table 1 the amplitude distribution functions $H(A, r)$
for the Anderson model in 2d can be rescaled to obtain a data collapse
similar to Fig.~\ref{fig:2}.  The result is shown in Fig.~\ref{fig:4},
again for $w=10.0$ with PBC and $w=8.5$ with HWBC.  The numerical results
confirm the log-normal shape of the amplitude distribution $H(-\ln \vert
\psi \vert,r)$.

In summary, we find two major deviations from the standard SPST for the 
Anderson model in 2d:  (i) The {\it logarithmic} $\lambda(r)$ dependence 
indicates that an asymptotic value $\lambda \equiv \gamma^{-1}$ (assumed in
the SPST) might be ill defined, since the logarithmic function does not
converge.  (ii) The fluctuations still follow the SPST hypothesis 
Eq.~(\ref{eq:flu2}), if $\lambda$ is replaced by the $r$-dependent 
localization length $\lambda(r)$ in Eq.~(\ref{lambdar}), and an additional
prefactor $\sigma/2$ is inserted.  But since this prefactor is smaller
than one, the fluctuations in the Anderson model in 2d are smaller than
those predicted by the standard SPST hypothesis. 

In order to find out, if these two types of deviations from the 
standard SPST are a generic feature of any system in 2d, we have 
determined the amplitude distributions and the values of $\lambda$ 
and $\sigma$ for electronic eigenfunctions on percolation clusters at 
the critical concentration, which we had studied in a previous 
work.\cite{kantelhardt-97}  Percolation\cite{bundestauffer} is a 
standard model for structurally disordered solids.  For site 
percolation, the Schr\"odinger equation in tight-binding approximation 
is identical to Eq.~(\ref{eq:TB}) with $\epsilon_n = 0$ for occupied 
sites (concentration $p$), and $\epsilon_n = \infty$ for unoccupied 
sites (concentration $1-p$).\cite{kirkpatrick-72}  The electronic wave 
functions are known to be localized for percolation clusters at the 
critical concentration $p = p_{\rm c}$ for any embedding dimension.
We found that the amplitude distributions at given topological distances 
from the localization center again obey the log-normal distribution 
(\ref{eq:pvonA}) for clusters in 2d as well as in 3d.  A data collapse 
is obtained with constant parameter $\lambda$, and no logarithmic 
dependence of $\lambda$ on the distance is observed.  Thus, the first 
deviation from the standard SPST that we observed in the 2d Anderson 
model does not show up for the percolation model.  Nevertheless, the 
second deviation, the modified value of the fluctuation parameter 
$\sigma$, also occurs in percolation.  But in contrast to the reduced 
value $\sigma = 1.3$ for the Anderson model in 2d (compared with the 
standard SPST value $\sigma=2$), we find increased values $\sigma 
\approx 2.6$ and 3.0 for percolation in 2d and 3d, respectively, which 
are still smaller than $\sigma=4$ for the Lloyd model.  We conclude that
a model dependent second parameter $\sigma$ is required to describe the
scaling behavior of the amplitude fluctuations.

\section{Analytical results for the mean wave function}

Since we can very well approximate the distribution of eigenfunction
amplitudes $\vert \psi \vert$ at large distances $r \gg \lambda$ by 
the log-normal distribution (\ref{eq:pvonA}), we can directly 
calculate the decay of the average wave function $\langle \vert \psi(r) 
\vert \rangle$ for large $r$ by simple integration.  We can also 
estimate the way this average, which we denote by $\Psi_N(r)$, depends 
on the number of averaged configurations $N$.  In particular, we can 
obtain $\Psi_N(r)$ in the asymptotic regime which is not accessible 
numerically.  The calculation is valid for Anderson wave functions in 
1d and 2d (as well as for the percolation model), since the same 
log-normal distribution $H_{\rm ln}(A,r)$, applies to all cases and 
only the two parameters $\lambda$ and $\sigma$ have to be adapted.  
Note, that the $r$-dependent localization length $\lambda(r)$ from 
Fig.~\ref{fig:2} has to be used for the Anderson model in 2d, while 
$\lambda$ is constant otherwise.

In order to describe the localization behavior properly, we are led to
the ansatz
\begin{equation} \ln \Psi_N(r) \sim -r^{d_\Psi}, \label{eq:dpsi}
\end{equation}
where $d_\Psi$ is an effective localization exponent.  If we average 
over all configurations, the resulting quantity $\Psi_\infty(r)$ is 
related to the distribution $H(A,r)$ by
\begin{equation}  \Psi_\infty(r) = \langle \vert \psi(r) \vert 
\rangle = \int_0^\infty \exp(-A) H(A,r) \: dA
\label{eq:theotra1} \end{equation}
with $A \equiv -\ln \vert \psi \vert$.  Note that only positive values of
$A$ are possible because of our normalization $\psi_{n_0}=1$ for the
localization center (maximum) at site $n_0$.  In the preceeding Section
we have numerically shown that $H(A,r)$ can be well approximated by
the log-normal distribution $H_{\rm ln}(A,r)$ for large distances $r$ 
from the localization center, where the amplitudes $\vert \psi \vert$ 
are small and $A \equiv -\ln \vert \psi \vert$ large.  Thus, we insert 
$H_{\rm ln}(A,r)$ for $H(A,r)$ in Eq.~(\ref{eq:theotra1}).  However, the 
exact shape of $H(A,r)$ for small distances, corresponding to large 
amplitudes, is not known exactly.  Hence, our result will be exact only 
for large values of $r$.  We further have to assume that the number of 
configurations $N$ is finite, so that the extreme tails of the distribution 
$H(A,r)$, i.~e. the very rare events possibly involving small $A$, are not 
important.  Our treatment includes a description of the typical average
over one configuration (one eigenfunction), which corresponds to $N=1$.

For a finite number $N$ of configurations, the total number of sites at 
distance $r$ from the localization center is identical to $N N_r$ with 
$N_r = a r^{d-1}$ and $a = 2$ ($2 \pi$) for the Anderson model in 1d (2d). 
Clearly, those values of $\vert\psi\vert$ with a too small probability, 
i.~e. probability smaller than $1/(N\: N_r)$, are unlikely to occur in 
$N$ typical configurations.  Thus, we must limit the distribution 
$H_{\rm ln}(A,r)$ by cutting off the area $1/(2 N\: N_r)$ at each of the 
two tails.  According to Eq.~(\ref{eq:pvonA}) the area below a lower 
cutoff $A_{\rm min}$ is given by
\begin{equation} \int_{-\infty}^{A_{\rm min}} H_{\rm ln}(A,r) \: dA
= {1 \over 2} \left\{1 + {\rm erf} \left[ \left( A_{\rm min} -
{r \over \lambda} \right) \sqrt{\lambda \over \sigma r} \right] \right\},
\label{eq:Amin1} \end{equation}
where ${\rm erf}(x) \equiv 2\pi^{-1/2} \int_0^x \exp(-t^2) \, dt$ is the
error function.  Equating this area to $1/(2 N\: N_r) = (2 N a 
r^{d-1})^{-1} $ and using the inverse error function ${\rm erfinv}(x)$ 
we obtain the lower cutoff value
\begin{equation} A_{\rm min} = {\rm max} \left\{ 0, {r \over \lambda} 
- \sqrt{\sigma r \over \lambda} {\rm erfinv} \left( 1 - {1 \over N a 
r^{d-1}} \right) \right\}, \label{eq:Amin2} \end{equation}
which replaces the lower integration bound in Eq.~(\ref{eq:theotra1}). 
Hence, for finite $N$, Eq.~(\ref{eq:theotra1}) becomes
\begin{equation} \Psi_N(r) = \int_{A_{\rm min}(r,N)}^\infty \exp(-A)
H_{\rm ln}(A,r) \: dA. \label{eq:theotra2} \end{equation}
The integration can be performed straightforwardly and gives
\begin{eqnarray} \Psi_N(r) = {1 \over 2}\exp\left[- \left( 1- 
{\sigma \over 4} \right) {r \over \lambda} \right] \times \nonumber \\ 
\left\{ 1-{\rm erf}\left[\sqrt{ {\sigma r\over 4\lambda} }-
{\rm erfinv} \left( 1 - {1 \over N a r^{d-1}} \right) \right] \right\}. 
\label{eq:psithe} \end{eqnarray}
The first term of this equation seems to indicate a regular exponential 
decay, if the second term could be disregarded and $\lambda$ was 
constant.  But, indeed, the second term, involving the error function 
as well as a complicated dependence on the number of configurations $N$, 
cannot be disregarded.  Thus, the average wave function $\Psi_N(r)$ 
depends explicitly on $N$ and exhibits significant deviations from the 
exponential from.  

\begin{figure} \centering
\epsfxsize8.cm\epsfbox{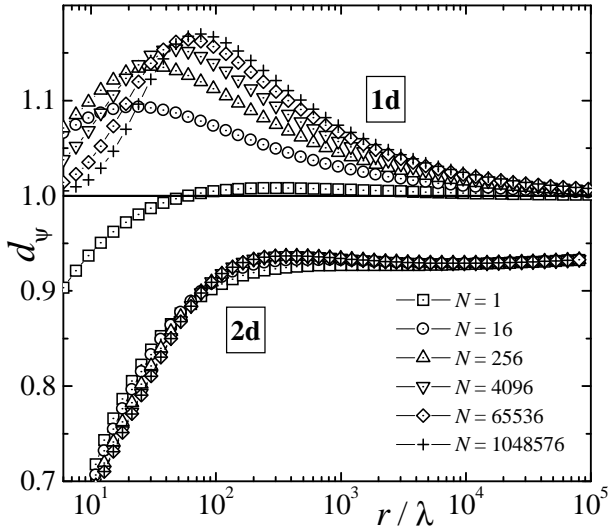}
\parbox{8.5cm}{\caption[]{\small
The theoretical effective localization exponents $d_\Psi$ determined
by fits of the ansatz \protect(\ref{eq:dpsi}) to the analytical result
\protect(\ref{eq:psithe}) are shown versus $r/\lambda$ for several 
numbers of averaged configurations $N$ (see legend).  For the 1d results
(upper curves) the parameters $\lambda=105$ and $\sigma=2$ have been 
used, while $\lambda(r)/0.89 = \ln(r/1.2)$ (from 
Fig.~\protect\ref{fig:3}) and $\sigma=1.3$ for the 2d case.  In the 1d 
case we find strong dependence on $N$ and $d_\Psi > 1$, indicating 
superlocalization, while $d_\Psi < 1$ indicating sublocalization is
observed in the 2d case. }
\label{fig:6}} \end{figure}

Equation~(\ref{eq:psithe}) is supposed to be rigorous for sufficiently 
large $r$ values, where the distribution of $\vert\psi\vert$ is exactly 
described by the log-normal distribution $H_{\rm ln}(A,r)$.  The decay of 
$\Psi_N(r)$ can be approximated by Eq.~(\ref{eq:dpsi}) with an effective 
localization exponent $d_\Psi$, which depends on $r$ and $N$.  In 
Fig.~\ref{fig:6}, we trace $d_\Psi$ as a function of $r/\lambda$ for 
several values of $N$ using the parameters $\lambda(r)$ and $\sigma$ 
corresponding to the Anderson model in 1d and 2d.  We determined 
the values of $d_\Psi$ by local fits of Eq.~(\ref{eq:dpsi}) to 
Eq.~(\ref{eq:psithe}).  Surprisingly, the behavior is very different in 
1d and 2d.  

In 1d, we find that the effective exponent $d_\Psi$ in the exponential 
ansatz (\ref{eq:dpsi}) is larger than one except for the typical average
($N=1$) which we will discuss later.  The localization is stronger than 
the usual exponential localization.  Thus, this regime can be described
by the term ``superlocalization''.  Furthermore, the values of $d_\Psi$ 
strongly depend on the number of configurations $N$.  We observe a local 
maximum in the dependence of $d_\Psi$ on $r$ for $r \approx 10 \lambda$ to 
$100 \lambda$.  While the height of the maximum increases with increasing
$N$, indicating stronger superlocalization for an increasing number of 
configurations, the position of the maximum is shifted towards larger $r$
values.  Only in the asymptotic limit for extremely large $r$ values, 
$r \approx 10^5 \lambda$, the effective localization exponent $d_\Psi$ 
converges to one, yielding the theoretically predicted simple exponential 
behavior.  

The typical average $\Psi_1(r)$, describing the behavior of one typical 
eigenfunction, can be deduced by setting $N=1$.  As can be seen in 
Fig.~\ref{fig:6}, the exponent $d_\Psi$ is smaller than one for
intermediate distances $r$ from the localization center (for $r<60 
\lambda$).  For larger $r$ it is very close to, but slightly above one. 
Practically, this behavior can no longer be distinguished from simple
exponential localization for $r>50 \lambda$.

In 2d, however, the localization behavior appears to be remarkably 
different.  As can be seen in Fig.~\ref{fig:6}, we observe hardly any 
$N$-dependence of the effective localization exponent $d_\Psi$, although 
the absolute value of the average wave function does slightly depend on 
$N$.  The local maximum of $d_\Psi$ is so weak, that it will be practically 
invisible.  The most important difference to the localization behavior in 
1d, though, is the small value of $d_\Psi$.  The effective localization 
exponent is smaller than one not only for intermediate $r$ values, but even 
remains below one asymptotically up to at least $r = 10^5 \lambda$.  Thus 
simple exponential behavior of the average wave function can apparently 
{\it not} be reached for any practical system size.  This is mainly due 
to the logarithmic dependence of the localization length $\lambda$ on 
$r$, that we have reported in Fig.~\ref{fig:3} and inserted into
Eq.~(\ref{eq:psithe}) to derive $d_\Psi$.  Consequently, this regime
is not characterized by superlocalization as in 1d.  We rather find a 
stretched exponential decay, so that the term ``sublocalization'' seems
to be appropriate, because the localization behavior is weaker than 
simple exponential.

Equation~(\ref{eq:psithe}) is not valid for small $r$ values where
the real distribution $H(A,r)$ deviates from the log-normal form 
$H_{\rm ln}(A,r)$.  Thus, deviations occur for $r$ approximately smaller 
than $10 \lambda$.  Further deviations occur for intermediate $r$ and 
large $N$ values when the tails of the distribution become more 
important.  Then the lower integration limit $A_{\rm min}(r, N)$ in 
Eq.~(\ref{eq:theotra2}) tends to zero and the distribution of rare 
events of wave function amplitudes $\vert\psi\vert$ close to the maximum 
(corresponding to $A \equiv -\ln \vert\psi\vert$ close to zero) becomes 
important.  

Despite of this, the log-normal approximation $H_{\rm ln}(A,r)$ can
be used for obtaining a qualitative picture of $\Psi_N(r)$ also at small
$r$ values, where $A_{\rm min}(r, N)=0$ in Eq.~(\ref{eq:Amin2}).  We
introduce the cutoff at $A=0$, since no amplitude $\psi_n$ at distance
$r>0$ from the localization center $n_0$ can be larger than 
$\vert\psi_{n_0}\vert = 1$, and thus $A\equiv-\ln \vert\psi_n\vert > 0$.  
It can be shown numerically, that the integration of
Eq.~(\ref{eq:theotra1}) yields an effective localization exponent
much smaller than one, $d_\Psi \approx 0.5$, for small $r$ in the ansatz
Eq.~(\ref{eq:dpsi}).  Thus, the decay is weaker than exponential in the
small-$r$ regime, and we expect a regime of sublocalization.  With
increasing $N$, $A_{\rm min}(r, N)$ in Eq.~(\ref{eq:theotra2}) decreases
reaching zero also for larger $r$ values.  Thus, the small-$r$ regime 
with $A_{\rm min}(r, N) = 0$ becomes dominant for averages over a large 
number of configurations.  Since the lower integration limit becomes zero
in Eq.~(\ref{eq:theotra2}) for the small-$r$ regime, it is evident that 
$\Psi_N(r)$ does not depend on $N$.

\section{Numerical results for the mean wave function}

\subsection{Averaging procedure}

Next we determine numerically the way the average amplitude values
$\vert\psi(r)\vert$ decay with increasing distance $r$ from the
localization center.  The necessary averaging procedure consists of
three steps.\cite{kantelhardt-97,bunde-95,bunde-99}

\begin{figure} \centering
\epsfxsize7.8cm\epsfbox{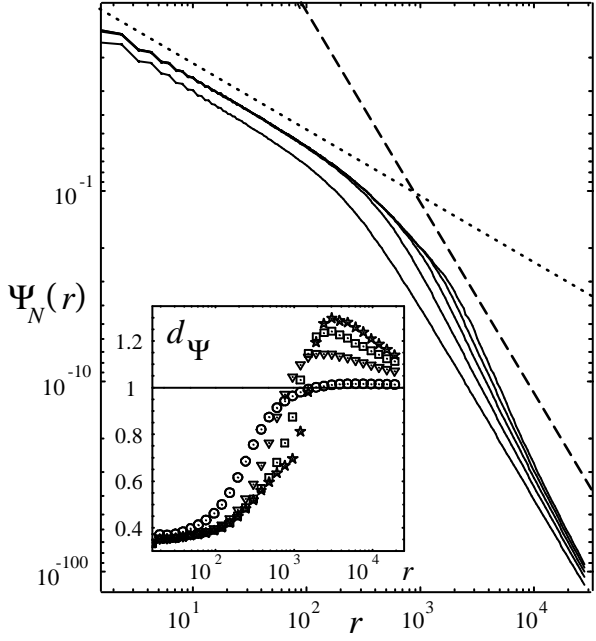}
\parbox{8.5cm}{\caption[]{\small
The decrease of the typical average amplitudes for $N$ configurations, 
$\Psi_N(r)$, versus $r$ for electronic wave functions in the Anderson 
model in 1d for $w=1.0$ and $N=1, 16, 256, 4096$, and 65536 (from the 
bottom to the top).  For the plot, more than $10^6$ eigenfunctions 
have been calculated on chains of lengths $L=30000$.  The straight 
lines with the slopes $d_\Psi=1$ (dashed) and $d_\Psi=0.4$ (dotted) 
are shown for comparison.  In the inset, the effective local exponents 
$d_\Psi$ determined numerically from the slopes of the curves are 
shown versus $r$.  The symbols correspond to the effective
numbers of  configurations, $N=1$ (circles), 16 (triangles),
256 (squares), and 4096 (stars).}
\label{fig:7}} \end{figure}

1. In the first step we average, for each eigenfunction 
$\psi_n^{(\nu)}$ separately, the values of $\vert\psi_n^{(\nu)}\vert$
for all sites $n$ at given distances $r$ from the localization center. 
The resulting function $\psi^{(\nu)}(r)$ characterizes the spatial
decrease of the $\nu$th eigenfunction. 

2. In the second step, for obtaining the mean spatial decrease of $N$
eigenfunctions, we average $\psi^{(\nu)}(r)$ over $N$ configurations
(eigenfunctions),
\begin{equation} \langle \psi(r) \rangle_N \equiv {1\over N}
\sum_{\nu=1}^N \psi^{(\nu)}(r). \label{eq:mittel2} \end{equation}
If the system is not selfaveraging (which we find is the case here),
the resulting values for $\langle \psi(r) \rangle_N$ will
depend on the special set of $N$ configurations considered and will
fluctuate from set to set.

3. To obtain the {\it typical} value of $\langle\psi(r)\rangle_N$, we
are led, in the third step, to the logarithmic average over many sets
of $N$ configurations, i.~e. we average $\ln \langle\psi(r)\rangle_N$
over many sets of $N$ eigenfunctions,\cite{kantelhardt-97,bunde-95,bunde-99}
\begin{equation} \Psi_N(r) \equiv \exp \langle \ln\langle\psi(r)\rangle_N
\rangle. \label{eq:mittel3} \end{equation}
In the following, we shall discuss exclusively this ``typical average
over $N$ configurations'' defined by Eq.~(\ref{eq:mittel3}).

\begin{figure} \centering
\epsfxsize7.8cm\epsfbox{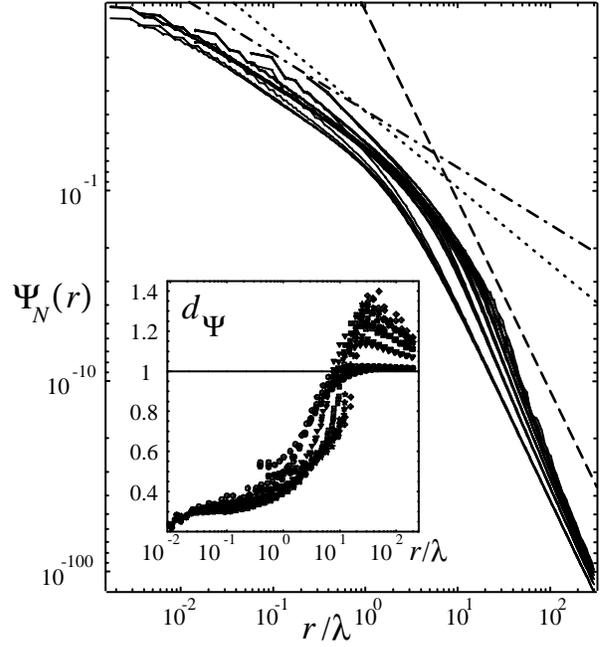}
\parbox{8.5cm}{\caption[]{\small
The decrease of the typical average amplitudes for $N$ configurations,
$\Psi_N(r)$, versus $r/\lambda$ for electronic wave functions in the
Anderson model in 1d for $N=1, 16, 256, 4096$, and 65536 (from the bottom
to the top) and five different strengths of disorder ($w=0.35, 0.5,
1.0, 2.0$, and 4.0).  For the total number of configurations, the
values of $\lambda(w)$, and the system sizes, see Fig.~2.  The straight
lines with the slopes $d_\Psi=1$ (dashed), $d_\Psi=0.4$ (dotted), and 
$d_\Psi=0.3$ (dot-dashed) are
shown for comparison.  In the insets, the effective local exponents
$d_\Psi$ determined numerically from the slopes of the curves, are
shown versus $r/\lambda$.  The symbols correspond to the effective
numbers of configurations $N=1$ (circles), 16 (triangles),
256 (squares), 4096 (stars), and 65536 (diamonds).}
\label{fig:8}} \end{figure}

The special case $N=1$ in the definition corresponds to the ``typical
average'' over one configuration.  Since $\langle \psi(r)
\rangle_1 \equiv \psi^{(\nu)}(r)$ in Eq.~(\ref{eq:mittel2}), the second
step in the averaging procedure is dropped, and the values from each
configuration are averaged logarithmically.

\subsection{Results in 1d}

Figure~\ref{fig:7} shows $\Psi_N(r)$ for the electronic
wave functions in 1d for several $N$ values.  The eigenfunctions are
calculated for states in the band center ($E \approx 0$) by the iteration
method,\cite{roman-87} and the disorder parameter was chosen to be $w=1.0$.
As can be seen in the Figure, the actual values of $\Psi_N(r)$
depend significantly on the number of configurations $N$ included in the
averaging procedure.  The local slopes of the curves yield the effective
exponent $d_\Psi$ in our ansatz Eq.~(\ref{eq:dpsi}) which are shown in
the inset of Fig.~\ref{fig:7}.  One can see that the decay of $\Psi_N(r)$
is not simple exponential ($d_\Psi=1$), but rather two different
localization regimes with evidently different effective exponents $d_\Psi$
can be distinguished.

\begin{figure} \centering
\epsfxsize7.8cm\epsfbox{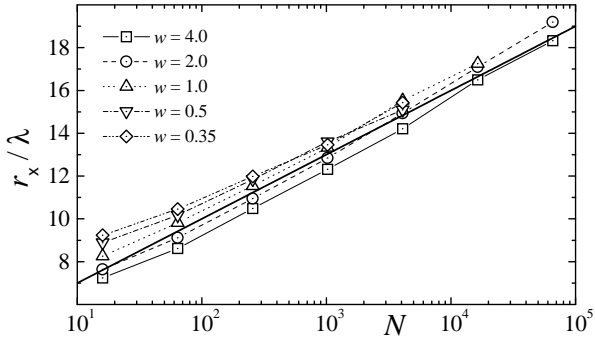}
\parbox{8.5cm}{\caption[]{\small
The crossover distances $r_\times(N)/\lambda$ for electronic wave
functions in the Anderson model in 1d are shown versus $N$ in a
semi-logarithmic plot.  The values of $r_\times(N)$ are determined as 
the lengths $r$, where the effective localization exponent $d_\Psi$ of 
$\Psi_N(r)$ intercepts with 1.  The symbols correspond to the different 
disorder strengths, $w=0.35$ (diamonds), 0.5 (stars), 1.0 (squares), 
2.0 (triangles), and 4.0 (circles).  The straight line
$r_\times(N)/\lambda = 3 \log_{10} N + 4$ is shown for comparison.}  
\label{fig:9}} \end{figure}

In the first localization regime, where $\Psi_N(r)$ decays roughly by one
order of magnitude, we find a stretched exponential decay (sublocalization)
with an effective exponent $d_\Psi \cong 0.4$ for all values of $N$, which
is consistent with the result obtained in the previous
Section.  As expected, this sublocalization regime expands as the number
of configurations $N$ increases.

In the second regime, for large $r$ values, self-averaging fails and
$\Psi_N(r)$ depends explicitly on $N$.  In the beginning of this
regime, $d_\Psi$ is considerably larger than $1$, corresponding to a
faster than exponential decay (superlocalization).  Only for extremely
large $r$ values the simple exponential decay ($d_\Psi=1$) is reached.
This behavior agrees nicely with the analytical results obtained in
the previous Section, as can be seen by comparing the inset of
Fig.~\ref{fig:7} with the analytical values of the effective localization
exponent $d_\Psi$ shown in Fig.~\ref{fig:6}.  Note that very large $r$
values cannot be reached in the numerical calculations.  Thus, the
theoretical description, that becomes exact asymptotically for large
$r$, is complementary to the numerics in this sense.

The localization behavior is similar for other strengths of
disorder $w$.  This can be seen in Fig.~\ref{fig:8}, where the
data for five different $w$ has been included.  Due to the scaling of
the localization length, $\lambda = 105/w^2$, the data can be scaled
onto master curves for each number of configurations $N$.  For the
second regime, where superlocalization occurs, we get an excellent
data collapse.  In the first regime, where sublocalization occurs,
the data collapse is less accurate, because the localization exponent
$d_\Psi$ depends on the disorder strength $w$.  We obtain
$d_\Psi \approx 0.3$ for small disorder ($w=0.5$) and $d_\Psi \approx
0.4$ for larger disorder ($w=2.0$).

The distance $r_\times$ characterizing the crossover from the first
($r<r_\times$) to the second ($r>r_\times$) localization regime depends
on the number of configurations $N$.  Since $d_\Psi<1$ in the
sublocalization regime (for $r<r_\times$) and $d_\Psi>1$ in the
superlocalization regime (for $r>r_\times$), the values of $r_\times$
can be determined as the lengths $r$ where $d_\Psi$ intercepts with
unity.  Figure \ref{fig:9} shows $r_\times(N)/\lambda$ versus
$\log N$.  It is obvious that $r_\times(N)$ increases approximately
logarithmically with $N$, $r_\times(N)/\lambda \approx 3 \log_{10} N +
4$.

\subsection{Results in 2d}

\begin{figure} \centering
\epsfxsize8.cm\epsfbox{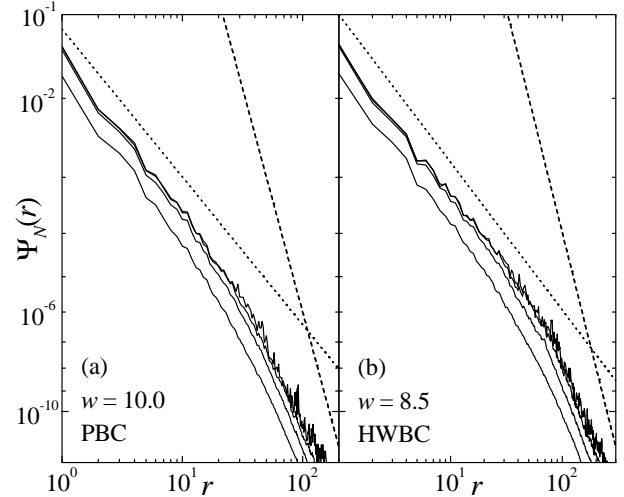}
\parbox{8.5cm}{\caption[]{\small
The decrease of $\Psi_N(r)$ versus $r$ for electronic wave 
functions in the Anderson model in 2d for $N=1$, 8, 64, and 512
(from the bottom to the top).  For (a) $w=8.5$ and HWBC were chosen,
while $w=10.0$ and PBC in (b).  For each part of the figure,
$\approx 1000$ eigenfunctions with $E \approx 0$ were calculated 
using the Lanczos algorithm on a $300 \times 300$ lattice.  The
straight dashed and dotted lines have the slopes $d_\Psi=1$ and 
$d_\Psi=0.33$, respectively, and are shown for comparison.}
\label{fig:10}} \end{figure}

Figure~\ref{fig:10} shows our results for the mean amplitudes
$\Psi_N(r)$ [defined in Eqs.~(\ref{eq:mittel2}) and (\ref{eq:mittel3})]
for the Anderson model in 2d.  We consider the same two disorder
strengths and types of boundary conditions as in Figs.~\ref{fig:3} and
\ref{fig:4} and four values of $N$. The Figure shows clearly that at
all distances from the localization center stretched exponentials
(sublocalization) occur, which is fully consistent with the theoretical
results obtained in Section IV (see especially Fig.~\ref{fig:6}).
In addition, there is practically no dependence of the effective
localization exponent, obtained from the local slopes in
Fig.~\ref{fig:10}, on the number of configurations $N$ taken into
account in the averaging procedure.  The mean wave function amplitudes
are just slightly shifted to larger values with increasing $N$.

In the small-$r$ regime, which cannot exactly be described by the
theoretical study in Section IV, we find strong sublocalization
characterized by $d_\Psi \approx 0.33$, similar to our results in
1d where we also obtained values between 0.3 and 0.4 (see
Figs.~\ref{fig:8} and \ref{fig:9}).  This regime ranges over
approximately two orders of magnitude in $r$ from the site neighboring
the center of localization ($r=1$) to $r_\times(N)$, which is $> 100$
for large $N$ in the two examples considered.  In this first regime
the mean wave function amplitudes decay to values of $\approx 10^{-5}$.
Thus, this regime is most relevant for most applications.

The second regime for large distances $r$ from the localization
center ($r > r_\times$) can be seen only in a quite limited $r$
range, because of the limited system size and finite size effects
especially for PBC.  The effective localization exponents $d_\Psi$ in
this regime are definitely larger than in the first regime, but we
still obtain values remaining strictly below one.  Thus, in 2d, the
second regime is also characterized by sublocalization, in contrast to
the superlocalization observed in 1d.  This numerical result is in very
nice agreement with the prediction from the theoretical description
in Section IV.

\section{Summary}

In order to discuss the localization and fluctuation behavior of
localized eigenfunctions in the Anderson model we have studied the
histogram amplitude distribution function $H(A,r)$ which corresponds
to the probability density to find the amplitude $A \equiv
-\ln\vert\psi\vert$ at distance $r$ from the localization center.
We find that $H(A,r)$ quite exactly follows a log-normal shape for
Anderson wave functions in 1d and 2d.  For increasing distance from
the localization center, the log-normal fits become even better.
Hence, the distributions can be characterized by two parameters, the
localization length $\lambda$ and the fluctuation parameter $\sigma$
that is proportional to the width of the distributions.  While both
parameters exactly follow the predictions of the standard single
parameter scaling theory in 1d, deviations occur in higher dimensions.
The fluctuation parameter is smaller than the standard value of the
single parameter scaling theory for the Anderson model in 2d, in
contrast to a larger value for the percolation model.  Furthermore, the
localization length shows a {\it logarithmic} dependence on the distance
from the localization center for the Anderson model in 2d, which gives
an explanation for the contradicting values that have been published
before and casts doubts on the existence of a finite asymptotic
localization length.

Using the log-normal ansatz for the amplitude distributions, the decay
of the average eigenfunctions can be calculated analytically.  It is
remarkable that by this simple approach, the essential features of the
localization phenomenon, in 1d: sublocalization in the regime of small
distances from the localization center, crossover to superlocalized
behavior (that depends logarithmically on the number of configurations
$N$), and final approach to simple exponential behavior, and in 2d:
sublocalization in all regimes, are reproduced.

\subsection*{Acknowledgements}

We would like to thank Richard Berkovits for useful discussions and Lev
Deych for helpful comments.  Support from the Deutsche Forschungsgemeinschaft
(DFG) and the German Academic Exchange Service (Deutscher Akademischer
Austauschdienst, DAAD) is gratefully acknowledged.

\end{multicols}
\end{document}